\documentclass[11pt]{article} 
\usepackage{amssymb,nath}
\def\eqref#1{{\rm(\ref{#1})}}
\def\vf#1{\lnull\frac{\partial}{\partial #1}\rnull}

\newtheorem{remark}{Remark}
\newtheorem{example}{Example}

\newtheorem{proposition}{Proposition}

\newenvironment{proof}{\removelastskip\bigskip\par\noindent{\bf Proof. }}%
{\quad$\square$\bigskip\par}

\def\Lie{\mathop{\mathfrak L}\nolimits}
\def\bigmetric{{\mathbf g}}

\def\symm{\,}

\begin{document}
\title{On local equivalence problem of spacetimes with two orthogonally transitive commuting Killing fields}

\author{M. Marvan and O. Stol\'{\i}n
\\\footnotesize Mathematical Institute, Silesian University in Opava,
\\\footnotesize Na Rybn\'\i\v cku 1, 746 01 Czech Republic}
\maketitle

\abstract{\footnotesize
Considered is the problem of local equivalence of generic four-dimensional metrics possessing two commuting and orthogonally transitive Killing vector fields.
A sufficient set of eight differential invariants is explicitly constructed, among them four of first order and four of second order in terms of metric coefficients. In vacuum case the four first-order invariants suffice to distinguish generic metrics.
\\[2ex]
{\bf Keywords:} metric equivalence problem.
\\
{\bf MSC:} 83C20, 35Q75.}

\normalsize
\mathindent = 3pc

\section{Introduction}

The metric problem of equivalence is the problem to decide whether two pseudo-Riemannian spaces are locally isometric. This question is of importance in general relativity (\cite[Ch.~9]{S-K-M-H-H}), where isometric means gravitationally identical.

Existing classifying algorithms depend on computation of curvature invariants, which are components of Riemann curvature tensor and its covariant derivatives in a particular frame specified in the course of the algorithm. This is the central idea of Cartan's~\cite{C,Br} solution to the equivalence problem by the method of moving frames, and the Karlhede~\cite{Ka,Ka-M} algorithm derived from it.
Karlhede classification lies behind two projects of publicly available databases of exact solutions of Einstein equations~\cite{I-L,P-S-dI3}.
In vacuum case, the Riemann tensor can be replaced with its traceless part, the Weyl tensor.
Presently, the Cartan--Karlhede curvature invariants are the only invariants known to solve the metric equivalence problem in full generality.

The main difficulty connected with the Cartan--Karlhede algorithm lies with the size of the data generated at the $n$th iteration of the main loop. The number and complexity of generated curvature invariants grow quite rapidly with the order $n$ of covariant derivatives.
During the past two decades, much effort has been exerted to tight the bounds on~$n$, with recent limit values being collected in~\cite[Ch.~9, Table~9.1]{S-K-M-H-H}. The required $n$ can be as high as five, even though the upper limits are reached only in very special cases~\cite{Sk}.
Understandably, the higher $n$ is, the bigger portion of generated invariants are actually redundant even if algebraic identities are accounted for as in MacCallum and \AA man~\cite{M-A}.
This is the price paid for leaving no metrics aside.
From the computational perspective, one is naturally interested in solving the equivalence problem in terms of minimal possible number of invariants of lowest possible order.
In this paper we provide such a solution within the class $\mathfrak T$ of generic metrics that possess two commuting orthogonally transitive Killing vectors.
Our choice reflects the abundancy of exact solutions of class $\mathfrak T$ in general relativity, be it stationary axisymmetric~\cite{Ca3} or cylindrically symmetric spacetimes. The orthogonal transitivity is an additional simplification that not only occurs for all explicitly known stationary axisymmetric metrics~\cite[p.~294]{S-K-M-H-H}, but can be derived from much simpler property of invertibility~\cite{Pa,Ca2}.
A fairly large portion of explicitly known spacetime metrics belong to this class, mainly due to the fact that powerful generation techniques operate within~$\mathfrak T$ or its vacuum subclass. Reduction to a rather compact Ernst equation~\cite{Er} was the key to solution by the inverse scattering method of Belinsky and Zakharov~\cite{B-Z}, as well as a number of B\"acklund transformations, such as the Harrison transformation~\cite{H}, Hoenselaers--Kinnersley--Xanthopoulos transformation~\cite{H-K-X}, Kinnersley--Chitre transformation~\cite{Ki-Ch}, and Neugebauer transformation~\cite{N}.
The majority of these results have been extended to metrics with suitable energy-momentum tensors, notably electro-magnetism. For relations between different solution-generating techniques see Cosgrove~\cite{C3}. For a number of explicit solutions see~\cite{S-K-M-H-H,Ve}.
As shown by Kinnersley~\cite{Ki}, every vacuum Petrov type $D$ metric belongs to  $\mathfrak T$, the result having been extended to aligned electro-vacuum metrics with cosmological constant by Debever and McLenaghan~\cite{D-ML}.
On the other hand, metrics from the class $\mathfrak T$ can be of any Petrov type except~{\it III.}
Last but not least, Cosgrove~\cite{C1} almost solved the equivalence problem for vacuum metrics in this class.
It should be noted that metrics of class $\mathfrak T$ pass the Karlhede classification rather smoothly, requiring no more than two covariant derivatives to be taken.
The Riemann tensor itself being of order two, the computed invariants are of order up to four in terms of metric coefficients.
It may therefore come as a surprise that appropriately chosen invariants of first and second order suffice, at least for generic metrics.
Our goal is to demonstrate that.

The content of the paper is as follows. In Section~2 we review the transformations that preserve the class~$\mathfrak T$. Then we compute the number of lowest order invariants to show that four of them are of the first order.

In Section 3 we present the four independent invariants
$C_\rho,C_\chi,Q_\chi,Q_\gamma$ as well as an invariant frame $X,Y$
on~$J^1$. They provide us with a solution to the equivalence problem
rather immediately. It should be stressed that we consider only
generic metrics within our class. The exact meaning of ``generic''
is specified along the solution of the equivalence problem. In
particular, metrics are required to satisfy the condition that their
Killing algebra possesses a unique two-dimensional commutative
subalgebra. However, we leave open the problem of recognizing this
condition in terms of first-order invariants.

Our work was to a great extent inspired by Cosgrove's treatment~\cite{C1} of the subclass of vacuum space-times. Exploiting an auxiliary invariantly defined metric of constant negative curvature, Cosgrove~\cite{C1,C2} devised a method to find new vacuum solutions of Einstein equations without cosmological constant. The work~\cite{C1} solves the equivalence problem for empty spacetimes modulo transformations of the natural harmonic coordinate $r$ and its companion coordinate.

In Section 4 we attempt a comparison between our and Petrov/Karlhede classification. The Cosgrove invariant $Q_\gamma$ is rather immediately identified with imaginary part of $\Psi_2$, another, $C_\chi$, with the real part of $\Psi_2$ in case of Ricci-flat (i.e., vacuum) metrics. We were unable to identify the others.
The last section is devoted to examples and discussion.

Beyond the scope of this paper we leave investigation of the non-generic cases as well as extension to wider classes of metrics. A general classification of conditions that can be put on a two-dimensional Killing distribution and its orthogonal complement can be found in~\cite{CF-V,Bach} along with the description of general solution in nongeneric cases.

\section {The metric and the pseudogroup}

\subsection{The metric}

In this section we introduce the class~$\mathfrak T$ of pseudo-Riemannian metrics of interest in this paper and the pseudogroup of local diffeomorphisms they admit.
Let $\mathcal M$ be a four-dimensional manifold. We consider a non-degenerate metric $\bigmetric$ on $\mathcal M$ possessing a unique two-dimensional commutative algebra $\mathfrak K$ of Killing vectors. Let $\Xi$ denote the vector distribution on $\mathcal M$ generated by $\mathfrak K$.
Then $\Xi$ is two-dimensional, since no non-constant multiple of a non-zero Killing vector can be a Killing vector.
Let $\xi_1,\xi_2$ denote arbitrary generators of the algebra~$\mathfrak K$.
We assume additionally that the Killing vectors be {\it orthogonally transitive}~\cite{Pa,Ca2}, meaning that the distribution $\Xi^\bot$ generated by the covectors $\bigmetric(\xi_i,\cdot\,)$ is Frobenius integrable and transversal to $\Xi$. Hence the tangent spaces decompose as $T_a \mathcal M = \Xi_a \oplus \Xi_a^\bot$.
By Frobenius integrability, the manifold becomes locally foliated by a two-parameter family $\mathcal S$ of two-dimensional leaves of the distribution $\Xi^\bot$. For an arbitrarily chosen leaf $\mathcal S_0$, points of $\mathcal S_0$ parametrize the orbit space~$\mathcal M/\Xi$, at least locally. Note that global properties of the foliation are irrelevant to the local equivalence problem.

Metrics of class $\mathfrak T$ have a simple description in terms of {\it adapted coordinates} (see also Weyl--Lewis--Papapetrou coordinates below).
Denoting by $\mathcal S_0$ a fixed surface of the family $\mathcal S$, we arbitrarily choose coordinates $t^1,t^2$ in an open subset of $\mathcal S_0$ and extend them along the trajectories of the fields $\xi_k$ to functions $t^1,t^2$ in an open subset $U$ of $\mathcal M$.
Shrinking $U$ if necessary, we determine functions $u^1,u^2$ in $U$ by the requirement that $\xi_l u^k = \delta^k_l$ and
$\mathcal S_0 \cap U = \{u^1 = u^2 = 0\}$.
In the coordinate system $(t^1,t^2,u^1,u^2)$ we have $\xi_k = \partial / \partial u^k$, $k = 1,2$, while the metric $\bigmetric$ assumes the form
$$\numbered\label{gg}
\bigmetric = g_{ij}\,`d t^i\,`d t^j + h_{kl}\,`d u^k\,`d u^l,
$$
where $g_{ij} = g_{ji},\ h_{kl} = h_{lk}$ are functions on $\mathcal S$, i.e., depend on $t^1,t^2$ only.
The induced metric $\bigmetric|_{\mathcal S}$ on $\mathcal S$ is
$g = g_{ij}\,`d t^i\,`d t^j$.
Obviously,
$$
`det \bigmetric = `det g `det h \ne 0.
$$
The other leaves of the foliation $\mathcal S$ are $u^1 = `const_1$, $u^2 = `const_2$.
Since shifts along the trajectories of Killing vectors are isometries, the particular choice of the leaf $\mathcal S_0$ is irrelevant. Henceforth we drop the index $0$ and write simply $\mathcal S$ instead of~$\mathcal S_0$.

The metric $\bigmetric$ can be identified with a section of the second symmetric power $S^2 T^* \mathcal M \to \mathcal M$ of the cotangent bundle $T^* \mathcal M \to \mathcal M$.
The two-dimensional spaces $\Xi_a$, $a \in \mathcal S$, constitute a vector bundle $\Xi \to \mathcal S$.
Considering the decompositions $T_a \mathcal M = \Xi_a \times T_a \mathcal S$ for $a \in \mathcal S$,
the metric $g = \bigmetric|_{\mathcal S}$ can be identified with a section of $S^2 T^* \mathcal S \to \mathcal S$, while the remaining metric coefficients $h_{kl} = \bigmetric(\xi_k,\xi_l)$ parametrize sections of the bundle $S^2 \Xi^* \to \mathcal S$.
Thus, locally, metrics $\bigmetric$ are in a one-to-one correspondence with sections $(g,h)$ of the six-dimensional vector bundle $\pi : S^2 T^* \mathcal S \times S^2 \Xi^* \to \mathcal S$. This correspondence, given simply by~\eqref{gg} in adapted coordinates, is only valid under the condition that the distributions $\Xi$ and $\Xi^\bot$ are fixed.

\subsection{The pseudogroup}

Turning back to our classification problem, we need to know the corresponding local transformation pseudogroup acting on sections of the bundle $\pi$. Otherwise said, what are local diffeomorphisms $\Phi$ of $\mathcal M$ such that $\bigmetric' = \Phi^*\bigmetric$ is of the same symmetry properties as $\bigmetric$ (both have the same algebra $\mathfrak K$, hence the same generators $\xi_1,\xi_2$ in $\mathfrak K$, and induce one and the same decomposition $T_a \mathcal M = \Xi_a \oplus \Xi_a^\bot$)? More to the point, what they are in terms of $g$ and $h$?

Let $\mathfrak G$ denote the pseudogroup of local diffeomorphisms of the base $\mathcal S$ of $\pi$. These have a unique lift along the bundle $S^2 T^* \mathcal S \to \mathcal S$, given by $g' = \phi^*g$, $\phi \in \mathfrak G$. Next consider the standard representation of $`GL_2$ in the two-dimensional vector space of the algebra $\mathfrak K$. That is, given $(a^k_l) \in `GL_2$, let $\xi'_k = a^i_k \xi_i$. Then we have the induced representation in each $\Xi_a$, hence in each $S^2 \Xi^*_a$, namely $h'_{kl} = \bigmetric(\xi'_k,\xi'_l)
 = a^i_k a^j_l h_{ij}$. Thus we arrive at the product $\mathfrak G \times `GL_2$, henceforth referred to as the pseudogroup of transformations of the total space $S^2 \Xi^* \times S^2 T^* \mathcal S$ of $\pi$.

\begin{proposition}
Let two metrics $\bigmetric$ and $\bigmetric'$ on $\mathcal M$ possess a unique two-dimensional commutative algebra $\mathfrak K$ of Killing vectors, which induces one and the same decomposition $T_a \mathcal M = \Xi_a \oplus \Xi_a^\bot$, $a \in \mathcal M$. Then $\bigmetric' = \Phi^*\bigmetric$ locally with respect to a  diffeomorphism $\Phi$ of $\mathcal M$ if and only if the corresponding sections $(g,h)$ and $(g',h')$ of $\pi$ are locally equivalent with respect to the action of $\mathfrak G \times `GL_2$ given above.
\end{proposition}

\begin{proof}
To prove the `if' part, let $\phi$ denote a local diffeomorphism on $\mathcal S$ such that $g' = \phi^* g$ and let $(a^i_j) \in `GL_2$ be such that $h'_{kl} = a^i_k a^j_l h_{ij}$. Denoting by $\Phi$ the local diffeomorphism of $\mathcal M$ with coordinate description $u^k \circ \Phi = a^k_l u^l$, $t^i \circ \Phi = t^i \circ \phi$,
from \eqref{gg} we get $\Phi^* \bigmetric' = \bigmetric$.

To prove the converse, consider a local diffeomorphism $\Phi$ of $\mathcal M$ such that $\bigmetric' = \Phi^*\bigmetric$. Let $\xi_1,\xi_2$ be the basis of $\mathfrak K$.
Then $\Phi_* \xi_1,\Phi_* \xi_2$ are Killing vectors for $\bigmetric$ and belong to the unique commutative subalgebra $\mathfrak K$, therefore they are linear combinations of $\xi_1,\xi_2$ with constant coefficients. Denoting by $\alpha$ the local diffeomorphism of $\mathcal M$ with local description $u^k \circ \alpha = a^k_l u^l$, $(a_k^l) \in `GL_2$, $t^i \circ \alpha = t^i$, we have $\alpha_* \xi_k = \Phi_* \xi_k$.
Hence, $\Phi\circ\alpha^{-1}$ preserves both vectors $\xi_1,\xi_2$.
The coefficients $(a_k^l) \in `GL_2$ being uniquely determined, $\alpha$ can be regarded as a representation of $`GL_2$ by local diffeomorphisms of $\mathcal M$, equivalent to the standard representation of $`GL_2$ in each $\Xi_a$.

Now, $\Phi\alpha^{-1}\mathcal S = \Phi\mathcal S$ is another leaf of the family, i.e., $\beta\mathcal S$, where $\beta$ is a shift along the trajectories of $\xi_1,\xi_2$, i.e., $u^l \circ \beta = u^l + b^l$. Denoting $\phi := \beta^{-1} \circ \Phi \circ \alpha^{-1}$, we see that $\phi$ is a local diffeomorphism that preserves $\mathcal S$ as well as both vector fields $\xi_1,\xi_2$. Moreover, $\phi$ is uniquely determined by its restriction $\phi|_{\mathcal S}$, hence $\phi$ effectively belongs to the pseudogroup $\mathfrak G$. Thus the local diffeomorphism $\Phi$ can be decomposed as $\Phi = \beta \circ \phi \circ \alpha$
into an automorphism $\alpha \in `GL_2$, a local diffeomorphism of $\phi \in \mathfrak G$, and a shift.
However, shifts are isometries, hence $\bigmetric' = \Phi^*\bigmetric
 = \alpha^*\phi^*\beta^*\bigmetric = \alpha^*\phi^*\bigmetric$.
Thus we are left with the pseudogroup $\mathfrak G \times `GL_2$.
\end{proof}

The classification problem for metrics $\bigmetric$ thus reduces to identifying orbits of the action of $\mathfrak G \times `GL_2$ on sections of the bundle~$\pi$.
Following Lie's classical method, one has to find a sufficient number of independent scalar differential invariants of the action. These are functions on the jet prolongation $J^\infty \pi$ that are invariant with respect to the infinitesimal action of $\mathfrak G \times `GL_2$ on $J^\infty \pi$.
Thus the first steps to be done are to describe the action of the corresponding Lie algebra $\mathfrak g \times \mathfrak{gl}_2$ on $\pi$ and its extension to the jet prolongation~$J^\infty \pi$.

Starting from the description of elements of $\mathfrak G$ as diffeomorphisms $\phi$ given by
$t^k \circ \phi = \phi^k(t^1,t^2)$,
we see that $\mathfrak g$ acts by vector fields
$$
\numbered\label{gg2}
U_\psi = \psi^i \frac\partial{\partial t^i}
$$
with $\psi = (\psi^1(t^1,t^2),\psi^2(t^1,t^2))$ being an arbitrary couple of functions on $\mathcal S$.
To determine the extension $U_\psi^\pi$ of infinitesimal transformations \eqref{gg2} on $\pi$, we require that the Lie derivative
$U_\psi^\pi(g) = U_\psi^\pi(g_{ij}\,dt^i\,dt^j)$ be zero, i.e.,
$$
U_\psi^\pi g_{ij} 
 = -g_{is} \frac{\partial \psi^s}{\partial t^j}
 - g_{sj} \frac{\partial \psi^s}{\partial t^i}.
$$
Thus we have
$$
U_\psi^\pi = \psi^i \frac\partial{\partial t^i}
   - \sum_{i \le j} (g_{is} \frac{\partial \psi^s}{\partial t^j}
       + g_{sj} \frac{\partial \psi^s}{\partial t^i})
     \frac\partial{\partial g_{ij}}.
$$

The remaining infinitesimal generators are $V_a$ with $a \in \mathfrak{gl}_2$ corresponding to action of $`GL_2$ on components $h_{kl}$.
In terms of coordinates $u^k$ on $\mathcal M$ we have $V_a = a^k_l u^l \frac\partial{\partial u^k}$ with $a = (a^k_l) \in \mathfrak{gl}_2$ being an arbitrary constant $2 \times 2$ matrix, hence
$$
V_a^\pi h_{kl} = -h_{ks} a^s_l - h_{sl} a^s_k
$$
similarly as above.
Then
$$
\numbered \label{gl2}
V_a^\pi = -\sum_{k \le l} (h_{ks} a^s_l + h_{sl} a^s_k)
 \frac\partial{\partial h_{kl}},
$$
since components transversal to $\mathcal S$ are suppressed on $\pi$.

Next we consider the infinite-dimensional vector bundle $J^\infty \pi$. Recall that obvious coordinates along the fibres of $J^\infty \pi$ are the formal derivatives of $g_{ij}$ and $h_{kl}$ of all orders with respect to $t^1,t^2$, e.g., $g_{ij,k} = \frac{\partial g_{ij}}{\partial t^k}$, etc.
On $J^\infty \pi$ one has the usual total derivative
$$
D_k = \frac{\partial}{\partial t^k}
 + g_{ij,k} \frac{\partial}{\partial g_{ij}}
 + h_{ij,k} \frac{\partial}{\partial h_{ij}}
 + g_{ij,kl} \frac{\partial}{\partial g_{ij,l}}
 + h_{ij,kl} \frac{\partial}{\partial h_{ij,l}} + \dots
$$
By $\bar U_\psi := U_\psi^{J^\infty \pi}$ and $\bar V_a := V_a^{J^\infty \pi}$ we denote the well-known prolonged fields on $J^\infty \pi$, characterized by the defining relation $[\bar U_\psi, D_k] = -\frac{\partial\psi^s}{\partial t^k} D_s$.
For details and explicit formulas see~\cite{A-V-L,B-V-V,Ol,Ov}.
As is well known, scalar differential invariants can be identified with functions on $J^\infty \pi$ invariant with respect to the fields $\bar U_\psi$ and $\bar V_a$. These functions form a commutative associative $\mathbb R$-algebra, which can be thought of as algebra of functions on the orbit space $J^\infty \pi/(\mathfrak G \times `GL_2)$.

We finish this section with a proposition giving the number $N_r$ of independent $r$th order scalar differential invariants, as found by the classical method of Lie:

\begin{proposition} \label{Nr}
The dimension $N_r$ of the orbit space $J^r \pi/(\mathfrak G \times `GL_2)$ is given by the following table:
$$
\begin{array}{r|rrrrrr}
   r & 0 & 1 &  2 &  3 &  4 & \dots \\\hline
 N_r & 0 & 4 & 14 & 28 & 46 & \dots
\end{array}
$$
\end{proposition}

The proof reduces to routine counting the number of independent equations in the system $\bar U_\psi f = 0$, $\bar V_a f = 0$ on~$J^r \pi$.
Computations are considerably simpler when using the Weyl--Lewis--Papapetrou metric coefficients (see Sect.~\ref{WLP} below).

\begin{remark} \rm
\label{not}
A comment on a possible source of misunderstanding is due. Proposition~\ref{Nr} refers to scalar differential invariants as functions on the jet space $J^\infty \pi$. If such a function, say $F$, is evaluated for a particular metric $\bigmetric$, then it becomes a function on $\mathcal S$, which we shall denote as $F|_\bigmetric$ (formally $F|_\bigmetric = F \circ j^\infty \sigma_\bigmetric$, where $j^\infty$ denotes a jet prolongation of a section of the bundle~$\pi$ and $\sigma_\bigmetric$ is the section associated with~$\bigmetric$).
Analogous correspondences hold for other geometric objects such as forms and vector fields. Hence another interpretation of scalar differential invariants as functions on~$\mathcal S$.

Both interpretations are natural and indispensable. For instance, the order of an invariant can only be seen in the context of jet spaces, while the most natural way to construct an invariant consists in combining various invariant geometric constructions on $\mathcal S$, as demonstrated in the next section.
It is usually harmless to use one and the same notation with both interpretations and omit the symbol $|_\bigmetric$. However, one should bear in mind that independence of functions on~$\mathcal S$ is very different from that on~$J^\infty \pi$. The maximal number of independent functions is two on~$\mathcal S$, and unlimited on~$J^\infty \pi$.
\end{remark}

\section{The equivalence problem}

\subsection{First-order invariants}

It is the aim of this section to provide explicit formulas for the four independent scalar invariants of the first order predicted in Proposition~\ref{Nr}.
We utilize invariant geometric objects associated with
the metric $g$ and with the triple of functions $h_{11}, h_{12} = h_{21}, h_{22}$ on~$\mathcal S$.
Indices are raised and lowered with the metric $g$. Comma notation is used for partial derivatives taken with respect to coordinates $t^1,t^2$ on $\mathcal S$.

Geometric objects on $\mathcal S$ (functions, metrics) are obviously $\mathfrak G$-invariant. Therefore, of interest are $`GL_2$-invariant geometric objects on~$\mathcal S$. Of them, $`GL_2$-invariant functions are the scalar invariants sought.

The obviously $`GL_2$-invariant metric $g$ alone has no invariants
of order less than two (it has one invariant of order 2, the scalar
curvature~$R$, and $k - 1$ more independent invariants of order $k$
for each $k > 2$; see \.Zorawski~\cite{Zor}).

However, if $\alpha = \alpha_{ij}\,dt^i \symm dt^j$ is another $`GL_2$-invariant quadratic (meaning bilinear symmetric) form on $\mathcal S$, then the trace
$C_\alpha = \alpha^i_i = g^{ij}\alpha_{ij}$ is a $`GL_2$-invariant scalar.
Moreover, the volume forms $`vol_g = \sqrt{\left|`det g\right|}\,dt^1 \wedge dt^2$ and $`vol_\alpha = \sqrt{\left|`det \alpha\right|}\,dt^1 \wedge dt^2$ are a multiple of each other, hence
$Q_\alpha = `det\alpha/`det g$
is one more $`GL_2$-invariant scalar on $\mathcal S$.
Alternatively, invariants $C_\alpha$ and $Q_\alpha$ can be defined as coefficients of the characteristic polynomial $`det(\alpha - \lambda g)/`det g$ in $\lambda$.

Denote by $x$ the function
$$
x = `det h
$$
on $\mathcal S$. We already know that $x \ne 0$ everywhere since otherwise $\bigmetric$ would be degenerate at some point. It is easily checked that $x$ is invariant with respect to the subgroup $`SL_2 \subset `GL_2$, but not to its complement $\mathbb R \subset `GL_2$ consisting of scalings $x \mapsto c x$.
The covector $X = dx/x$ is then invariant with respect to the full $`GL_2$ since scalings cancel out. Let $\rho$ denote the degenerate quadratic form $(dx/x)^2 = (1/x^2) x_{,i} x_{,j}\,dt^i \symm dt^j$ on~$\mathcal S$.
The trace
$$
C_\rho = \frac{1}{x^2} g^{ij} x_{,i} x_{,j}
$$
is a $`GL_2$-invariant function on~$\mathcal S$, while
$Q_\rho$ is obviously zero.

Another $`GL_2$-invariant metric on $\mathcal S$ is
$\chi = \frac{dh_{11} \symm dh_{22} - dh_{12} \symm dh_{12}}{x}$
with components
$$
\chi_{ij} = \frac 1{2x} (\left|\begin{array}{cc} h_{11,i} & h_{12,j} \\ h_{21,i} & h_{22,j} \end{array}\right|
 + \left|\begin{array}{cc} h_{11,j} & h_{12,i} \\ h_{21,j} & h_{22,i} \end{array}\right|).
$$
Hence two more $`GL_2$-invariant functions on~$\mathcal S$, namely
$$
C_\chi = \frac 1x g^{ij}
 \left|\begin{array}{cc}
   h_{11,i} & h_{12,j} \\ h_{21,i} & h_{22,j}
 \end{array}\right|,
\qquad
Q_\chi = \frac{`det\chi}{`det g} .
$$

Next consider linear combinations $\chi + c \rho$, which
are $`GL_2$-invariant quadratic forms on $\mathcal S$ as well.
For the particular value $c = -\frac14$ we obtain the {\it Cosgrove form}
$$
\gamma = \chi - \frac14 \rho.
$$
This name reflect the fact that $\gamma$ coincides with \cite[Eq.~(2.3)]{C1}.
While $C_\gamma = C_\chi - \frac14 C_\rho$, the other invariant
$$ \numbered \label{Qgamma}
Q_\gamma = \frac{`det\gamma}{`det g}
 = \frac1{4 x^3 `det g}
\left|\begin{array}{lll}
  h_{11} & h_{12} & h_{22} \\
  h_{11,1} & h_{12,1} & h_{22,1} \\
  h_{11,2} & h_{12,2} & h_{22,2}
\end{array}\right|^2
$$
is, in general, functionally independent of the invariants $Q_\chi$, $C_\chi$, $C_\rho$ (on $J^1 \pi$; see Remark~\ref{not}). If so, then we choose $C_\rho, C_\chi, Q_\chi, Q_\gamma$ to be the {\it basic first-order invariants} all other first-order scalar invariants are functions of.

\subsection{Invariant frame}
\label{IF}

The invariant covector $dx/x$ on $\mathcal S$ can be turned into an invariant vector by rising its index. If done with the help of the metric $g$, the resulting vector $X$ has components
$$
X^i
 = g^{ij} \frac 1x \frac{\partial x}{\partial t^j}.
$$
If done with the help of the volume form $`vol_g$, the resulting vector $Y$ has components
$$
Y^i
 = \epsilon^{ij} \frac 1{x \sqrt{|`det g|}} \frac{\partial x}{\partial t^j},
$$
where $\epsilon^{11} = \epsilon^{22} = 0$, $\epsilon^{12} = -\epsilon^{21} = -1$.
It is perhaps worth noticing that
$X \ln x = C_\rho$ while $Y \ln x = 0$.

Since
$$
\left|\begin{array}{cc} X^1 & Y^1 \\ X^2 & Y^2 \end{array} \right|
 = \frac{C_\rho}{\sqrt{|`det g|}},
$$
vectors $X,Y$ form an invariant frame at each point of $\mathcal S$ where $C_\rho \ne 0$, i.e., where $X$ is not null.
Components of $g$ with respect to this frame are
$$
g(X,X) = C_\rho, \quad
g(X,Y) = 0, \quad
g(Y,Y) = \pm C_\rho,
$$
where the sign is that of $`det g$.
Components of $\rho$ are simply
$$
\rho(X,X) = C_\rho^2, \quad
\rho(X,Y) = 0, \quad
\rho(Y,Y) = 0.
$$
For components of $\chi$ we easily get
$$
\chi(X,X) = C_\rho C_\chi - 4 Q_\chi + 4 Q_\gamma, \\
\chi(Y,Y) = \pm 4 (Q_\chi - Q_\gamma), \\
\chi(X,X)\chi(Y,Y) - \chi(X,Y)^2 = \pm C_\rho^2 Q_\chi, \\
$$
hence
$$
\chi(X,Y) = \sqrt{\pm C_\rho^2 Q_\chi
 \mp 4 (Q_\chi - Q_\gamma) (C_\rho C_\chi - 4 Q_\chi + 4 Q_\gamma)}.
$$

\subsection{Second-order scalar invariants}
\label{soi}

Lie derivatives of invariant objects with respect to invariant vector fields such as $X,Y$ are invariant objects again. Moreover, the vector fields on $J^\infty \pi$ corresponding to $X,Y$ (see Remark~\ref{not}) are {\it invariant differentiations}, since they commute with the vector fields $\bar U_\psi$, $\bar V_a$ for all $\psi$ and $a$.
In particular, if $p$ is a first-order scalar invariant, then $X p$, $Y p$ are second-order scalar invariants.

Starting with the four basic first-order invariants $C_\rho, C_\chi, Q_\chi, Q_\gamma$, we obtain up to eight independent second-order invariants in this way.
As we shall demonstrate below, these suffice for solving the classification problem in generic case.

Perhaps the best-known scalar second-order invariant of $g$ is the scalar curvature $R$, which is the only independent invariant of second order (in terms of metric coefficients) on a two-dimensional manifold like~$\mathcal S$.

The invariant vector fields $X,Y$ are orthogonal with respect to $g$ but do not commute in general. The coefficients $a,b$ in the commutation relation $[X,Y] = a X + b Y$ are second-order scalar differential invariants again.

Possible ways to construct invariants of second order include finding $`GL_2$-invariant quadratic forms on $\mathcal S$, such as the Lie derivatives $\Lie_X \chi$ and $\Lie_Y \chi$, or symmetric products
$dp \odot dq$ for various choices of first-order scalar invariants $p,q$.
Another obvious candidate is the Ricci tensor $\mathrm{Ric}_{ij}$ (which, unlike the Ricci tensor $\mathbf{Ric}_{ij}$ of $\bigmetric$, does not necessarily vanish in case of vacuum metrics).
Next, associated with any smooth function $f$ on $\mathcal S$ is the hessian $`Hess f$, which is a quadratic form defined by $`Hess f(Z,Z') = i_{Z} \nabla_{Z'} df$, i.e.,
$$
`Hess_{ij} f = f_{,ij} - \Gamma^k_{ij} f_{,k}.
$$
Here $\nabla$ is the Levi-Civita connection associated with $g$ and $\Gamma^k_{ij}$ are the usual Christoffel symbols for~$\nabla$.
The trace of $`Hess f$ coincides with the Laplace--Beltrami operator
$$
\Delta f = g^{ij} `Hess_{ij} f.
$$
Note that the order of $`Hess f$ equals two plus the order od $f$.
To obtain a $`GL_2$-invariant of second order, $f$ must be of order zero.
Choosing $f$ to be the logarithm of the $`SL_2$-invariant $x$ of order zero, we get
$$
\nu_{ij} = `Hess_{ij} `ln x = \frac 1x `Hess_{ij} x + \rho_{ij}
$$
(note that scalings of $x$ cancel out).

Therefore, a number of second-order invariants result as $C_\alpha$, $Q_\alpha$ for $\alpha$ running through various linear combinations of $\mathrm{Ric}_{ij},\nu_{ij},\rho_{ij},\chi_{ij},(\Lie_X\chi)_{ij},(\Lie_Y\chi)_{ij}$.

On $\mathcal M$, we have of course also the geometric objects
associated with the full metric $\bigmetric$. They can be restricted
to $\mathcal S$, thereby providing another source of invariant
objects. Since at most 14 second-order invariants can be independent
(as functions on $J^2 \pi$), it should not come as a surprise that,
for instance, the restriction of the full Ricci tensor
$\mathbf{Ric}_{ij}$ is a linear combination of the Ricci tensor with
$\nu,\rho$ and~$\chi$:
$$  \numbered \label{RSij}
(\mathbf {Ric}|_{\mathcal S})_{ij} = \mathrm{Ric}_{ij}
 - \frac12 \nu_{ij}
 - \frac14 \rho_{ij}
 + \frac12 \chi_{ij}, \quad i,j = 1,\dots, 2.
$$
The remaining components of the full Ricci tensor can be written as
$$
\mathbf{Ric}_{2+i,2+j} =
 - \frac12 \Delta h_{ij}
 + \frac14 X h_{ij}
 - \frac12 C_\chi h_{ij} \quad i,j = 1,\dots, 2.
$$
The full scalar curvatures are then related by
$$ \numbered \label{mathbf R}
\mathbf R = R - C_\nu - \frac12 C_\rho + \frac12 C_\chi
$$
with $C_\nu = \Delta \ln x$.


\subsection{Weyl--Lewis--Papapetrou parametrization}
\label{WLP}

The four basic invariants $C_\rho, C_\chi, Q_\chi, Q_\gamma$ look considerably simpler when the coefficients $h_{ij}$ are expressed via the (three of four) Weyl--Lewis--Papapetrou~\cite{Lew} variables $r,s,w$ ($r,s$ nonzero) according to the identification
$$
h = \frac rs [(du^1 + w\,du^2)^2 \mp s^2\,(du^2)^2],
$$
i.e.,
\begin{equation}
h = \left(\begin{array}{cc}
\frac{r}{s} & \frac{r w}{s} \\
\frac{r w}{s} & \frac{r(w^2 \mp s^2 )}{s}
\end{array}\right)
\end{equation}
in matrix notation.
The minus sign corresponds to $`det h = -r^2 < 0$ and the plus sign to $`det h = r^2 > 0$. If $\xi_1$ is non-null, which can always be achieved, then
$h_{11} \ne 0$ and
$r = \sqrt{\left|`det h\right|}$,
$s = r/h11$, $w = h_{12}/h_{11}$.

Under this parametrization we have
$$
\numbered
\label{WLPgamma}
x = \mp r^2, \qquad
\rho = \frac4{r^2}\,dr^2, \qquad
\gamma 
 = \frac 1{s^2} (ds^2 \mp dw^2).
$$
Hence Cosgrove's~\cite{C1} crucial observation that $\gamma$ is a metric of constant curvature $-1$ if non\-degenerate, i.e., if $Q_\gamma \ne 0$.
Formulas
$$ \numbered \label{CQ-WLP}
C_\gamma = C_\chi - \frac14 C_\rho
 = -g^{ij} \frac{s_{,i} s_{,j} \pm w_{,i} w_{,j}} {s^2}, \qquad
Q_\gamma = \mp \frac 1{s^4 `det g} \left|
 \begin{array}{cc} s_{,1} & w_{,1} \\ s_{,2} & w_{,2} \end{array}
\right|^2
$$
will be needed in the sequel.
In case of Lorentz metrics we have $`det g `det h < 0$, hance $Q_\gamma \le 0$.

\begin{proposition} \label{Kg} 
If $Q_\gamma \ne 0$, then the Killing algebra of $\gamma$ coincides with the $\mathfrak{sl}_2$ component of the algebra $\mathfrak{gl}_2$ of infinitesimal transformations~\eqref{gl2} acting on $\Xi$.
\end{proposition}

\begin{proof}
Both algebras have one and the same set of generators
$$
\vf{w}, \quad
s \vf{s} + w\vf{w}, \quad
s w \vf{s} + \frac{1}{2} (w^2 \mp s^2) \vf{w}
$$
in terms of Weyl--Lewis--Papapetrou variables.
\end{proof}

Further simplification is achieved if taking into account that all two-dimensional metrics are conformally flat. Then $g$ can be cast in one of the following two forms:
\begin{equation}
\label{CF}
g = f \,(dt^1\,dt^1 + dt^2\,dt^2) \quad \text{or} \quad
g = 2f\,dt^1\,dt^2.
\end{equation}
The full set of Weyl--Lewis--Papapetrou variables contains~$f$ as the fourth variable.

It should be noted that when fixing the explicitly conformally flat form of the metrics \eqref{CF}, we thereby impose further restriction on the pseudogroup $\mathfrak G$, namely, that the local diffeomorphisms be conformal maps. Fortunately, the classification problem of general two-dimensional metrics with respect to diffeomorphisms and the classification problem of metrics~\eqref{CF} with respect to conformal maps are identical. Using the Weyl--Lewis--Papapetrou coordinates $r,s,w,f$ substantially simplifies computations on~$J^\infty \pi$.

\subsection{A solution to the equivalence problem}
\label{equiv}

In this section we show how to reconstuct a generic
metric~$\bigmetric$ up to isometry from the knowledge of functional
dependences between certain eight invariants computed
for~$\bigmetric$. This provides a solution to the local equivalence
problem as a particular case of the ``principle of $n$ invariants''
\cite[Ch.~7, \S~4.3]{A-V-L}. Let $p,q$ be two independent scalar
invariants, which $p,q$ remain functionally independent when
evaluated on a pseudo-Riemannian space $\bigmetric$ (genericity).
Then the metric $\bigmetric$ is locally uniquely determined by the
values of six components $\bigmetric_{ij}(p,q)$ with respect to
coordinates $p,q$; these components are themselves scalar
invariants. Now, two metrics $\bigmetric,\bigmetric'$ are locally
equivalent if and only if the corresponding invariants
$\bigmetric_{ij}(p,q), \bigmetric'_{ij}(p',q')$ have equal
expressions in terms of invariants $p,q$ and $p',q'$, respectively.
That expressions for $\bigmetric_{ij}(p,q)$ can be large is not a
problem, since it is $p,q$ which must be simple enough for the
procedure to be computationally tractable. In practice one usually
finds a simpler set of invariants than $\bigmetric_{ij}(p,q)$, which
is then suitable for storing in a database. Let $\bigmetric$ be
fixed in the sequel. In what follows we interpret invariants as
objects on $\mathcal S$ (cf. Remark~\ref{not}).

\begin{proposition} \label{EP}
Let $p,q$ be two
functionally independent invariants from the list $C_\rho,\ C_\chi,\
Q_\chi,\ Q_\gamma$ (on $\mathcal S$). Then knowing the values of the
remaining two invariants and the invariants $Xp, Yp, Xq, Yq$ in
terms of $p,q$ is sufficient for recovering the metric $\bigmetric
\in ~\mathfrak T$.
\end{proposition}

Recall that $Xp, Yp, Xq, Yq$ are just four of the numerous second-order invariants of $\bigmetric$ discussed in Section~\ref{soi}.

\begin{proof}
Once we know the values
$Xp, Yp, Xq, Yq$ in terms of $p,q$, we know the invariant frame $X,Y$ in terms of coordinates $p,q$:
$$
X = Xp\,\frac\partial{\partial p} + Xq\,\frac\partial{\partial q}, \qquad
Y = Yp\,\frac\partial{\partial p} + Yq\,\frac\partial{\partial q}.
$$
Knowing also the remaining two values of $C_\rho, C_\chi, Q_\chi, Q_\gamma$ in terms of coordinates $p,q$, we can apply formulas given at the end of Section~\ref{IF} to compute $g,\rho,\chi$, and hence also $\gamma$, in terms of coordinates $p,q$.
In particular, $g$ is fully and uniquely restored in this way.

To restore the remaining three components $h_{ij} = \mathfrak g(\xi_i,\xi_j)$, we
start with the determinant $x$, which is determined from $g(X,\adot) = dx/x$ up to a constant multiplier. In terms of Weyl--Lewis--Papapetrou parameters $r,w,s$, we have $x = \mp r^2$, hence $r$ is also restored up to a constant multiplier. To obtain $w,s$, we distinguish two cases.

If $Q_\gamma \ne 0$, then $\gamma$ is the Cosgrove metric of constant curvature, i.e., a space form.
As is well known, space forms are uniquely determined by their dimension and signature
(see. e.g., Eisenhart~\cite[Ch.~I, \S~10]{Ei}). Therefore, $\gamma$ is isometric to the form $\frac 1{s^2} (ds^2 \mp dw^2)$, meaning that $w,s$ can be restored up to isometry of $\gamma$ (isometries of $\gamma$ reflect the freedom of determination of $w$ and $s$ by Proposition~\ref{Kg}).
This last step has been elaborated by Cosgrove~\cite{C1} and reduced to solution of Appel equations.

If $Q_\gamma = 0$, then by \eqref{CQ-WLP} we have
$$
\left|
 \begin{array}{cc} s_{,1} & w_{,1} \\ s_{,2} & w_{,2} \end{array}
\right| = 0.
$$
Therefore $w,s$ are functionally dependent, i.e., $s = s(z), w = w(z)$ for some function $z$ on $\mathcal S$, and
$$
\gamma = \frac{s^{\prime 2} \mp w^{\prime 2}}{s^2} (dz)^2
$$
where prime denotes differentiation with respect to $z$.
Assuming $s$ non-constant, upon identification $z = s$ we have
$\gamma = (1 \mp w_s^2) (\frac{ds}{s})^2$,
which determines $w(s)$ up to an additive constant.
Assuming $s = `const$ and $w \ne `const$, the identification $z = w$ leads to
$\gamma = \mp \frac 1{s^2} (dw)^2$,
which determines $w$ up to an additive constant again.
Assuming $s = `const$, $w = `const$ would lead to $Q_\gamma = Q_\chi = 0$ and $C_\chi = \frac14 C_\rho$, contradicting the assumptions of the proposition.
\end{proof}

\subsection{Relation to $\Psi_2$}

The first-order invariants $C_\rho$, $C_\chi$, $Q_\chi$, $Q_\gamma$
are not necessarily invariant with respect to general
diffeomorphisms, since by very construction they depend on the
uniqueness property of the subalgebra~$\mathfrak K$. In this section
we relate $Q_\gamma$ to $`Im\Psi_2$ and $C_\chi$ to $`Re\Psi_2$,
where $\Psi_2$ is one of the Petrov invariants.

Consider the Weyl tensor $\mathbf C_{abcd} = \mathbf{Riem}_{abcd}
 - \mathbf{g}_{a[c}\mathbf{Ric}_{d]b} + \mathbf{g}_{b[c}\mathbf{Ric}_{d]a}
 + \frac{1}{3} \mathbf R \mathbf{g}_{a[c}\mathbf{g}_{d]b}$ in terms of adapted coordinates $t^1,t^2,u^1,u^2$.
Its nonzero components are
$$
\mathbf{C}_{1234} = \frac12 \sqrt{Q_\gamma `det h `det g}, \\
\mathbf{C}_{1212} = \frac16 (R + \frac12 \Delta\ln x + \frac14 C_\rho - C_\chi) `det g, \\
\mathbf{C}_{3434} = \frac16 (R + \frac12 \Delta\ln x + \frac14
C_\rho - C_\chi) `det h,
$$
components of the form $\mathbf{C}_{ikjl}$ ($i,j = 1,2$ and $k,l =
3,4$), which are too involved to be presented here, and components
obtained from these through the symmetries $\mathbf{C}_{abcd} =
-\mathbf{C}_{bacd} = -\mathbf{C}_{abdc} = \mathbf{C}_{cdab}$. Due to
its symmetries, the Weyl tensor can be thought of as acting on
bivectors. The corresponding eigenvalue problem lies in the heart of
the Petrov classification the Karlhede classification is a
refinement of. To determine the Petrov type of a Lorentz metric, one
examines the multiplicities of roots of the quartic $\Psi_0 x^4 - 4
\Psi_1 x^3 y + 6 \Psi_2 x^2 y^2 - 4 \Psi_3 x y^3 + \Psi_4 y^4$,
where $\Psi_0 = \mathbf{C}_{abcd}k^am^bk^cm^d$, $\Psi_1 =
\mathbf{C}_{abcd}k^al^bk^cm^d$, $\Psi_2 =
\mathbf{C}_{abcd}k^am^b\bar{m}^cl^d$, $\Psi_3 =
\mathbf{C}_{abcd}k^al^b\bar{m}^cl^d$, $\Psi_4 =
\mathbf{C}_{abcd}\bar{m}^al^bk\bar{m}^cl^d$ are calculated with
respect to an arbitrary complex null tetrad $(k,l,m,\bar{m})$.
Routine computation shows that $\Psi_1 = 0$ and $\Psi_3 = 0$ (hence
metrics in $\mathfrak T$ cannot be of Petrov type $\it III$) and
then the quartic in question becomes $\Psi_0 X^4 + 6 \Psi_2 X^2 Y^2
+ \Psi_4 Y^4$. A metric is algebraically special if and only if
$\Psi_0 \Psi_4 (9 \Psi_2^2 - \Psi_0 \Psi_4) = 0$. Moreover,
$\Psi_0\Psi_4$ and $\Psi_2$ are invariants with respect to general
diffeomorphisms. Of special interest in this paper is the invariant
$\Psi_2$ and its real and imaginary part. By routine computation
\begin{equation}
`Re\Psi_2 = \frac1{16} C_\chi - \frac18  R + \frac1{24} \mathbf R \label{RePsi2}
\\
`Im\Psi_2 = \frac14 \sqrt{-Q_\gamma}
\end{equation}
(recall that $Q_\gamma \le 0$ in case of Lorentz metrics). Thus,
$`Im\Psi_2$ is of the first order.

\subsection{Genericity}
\label{gener}

It is the aim of this section to review the genericity assumptions we made in the course of exposition and express them in terms of scalar invariants.
Here is the full list of the assumptions (additionally to $`det g \ne 0$ and $x = `det h \ne 0$, which follow from the nondegeneracy of $\bigmetric$):

\begin{enumerate}
\item The algebra of Killing vectors has a unique two-dimensional commutative subalgebra $\mathfrak K$ (Section~2);

\item $C_\rho \ne 0$ (Section~\ref{IF});

\item There are at least two functionally independent invariants among $C_\rho$, $C_\chi$, $Q_\chi$, $Q_\gamma$ (Proposition~\ref{EP}).
\end{enumerate}

The first of these assumptions, unlike the others, is not formulated
in terms of invariants. Moreover, it cannot be formulated in terms
of invariants that themselves depend on the uniqueness property of
$\mathfrak K$. Yet we can improve the situation with the help of a
classical result valid without restriction to the class $\mathfrak
T$.

\begin{proposition}
Let $\bigmetric$ be a general metric (possibly outside class $\mathfrak T$), for which there exist two functionally independent scalar invariants with respect to the general diffeomorphisms. Then the distribution generated by Killing vectors of $\bigmetric$ is of dimension at most two.
\end{proposition}

\begin{proof}
More general results are proved in Kerr~\cite{Ke1,Ke2}. A simple proof goes as follows.
Assume that $\xi_1,\xi_2,\xi_3$ are three Killing vectors of $\bigmetric$ such that the distribution $\Xi$ they generate is of dimension three. Let $p,q$ be scalar differential invariants. Then $dp$ vanishes on~$\Xi$, since $0 = \xi_i p = dp(\xi_i)$, and similarly $dq$ vanishes on~$\Xi$. But $\mathcal M$ has only one dimension more than that of $\Xi$, whence $dp \wedge dq = 0$, proving that invariants $p,q$ are functionally dependent.
\end{proof}

Turning back to metrics of class $\mathfrak T$, our $Q_\gamma$ is invariant with respect to general diffeomorphisms in view of results of the preceding section. The same is true about $C_\chi$ in case of vacuum Einstein metrics.

Now we are left with the question when there exists a third Killing vector lying in the distribution $\Xi$ generated by $\mathfrak K$.

\begin{proposition} \label{aKv}
The algebra of Killing vectors spanning the distribution $\Xi$ is three-dimensional if and only if coefficients $h_{ij}$ are constant multiples of each other and two-dimensional otherwise.
\end{proposition}

\begin{proof}
In the adapted coordinates we have
$\xi_1 = \partial/\partial u^1$, $\xi_2 = \partial/\partial u^2$,
$\xi_3 = f^1\,\partial/\partial u^1 + f^2\,\partial/\partial u^2$ for some functions~$f^i$ of $t^1,t^2,u^1,u^2$.
The Killing equations $\xi_3 \bigmetric = 0$ reduce to
\begin{eqns}
h_{ik} \frac{\partial f^k}{\partial t^j} = 0, \quad i,j = 1,2, \label{Ke1} \\
h_{ik} \frac{\partial f^k}{\partial u^j}
 + h_{kj} \frac{\partial f^k}{\partial u^i} = 0, \quad i,j = 1,2. \label{Ke2}
\end{eqns}
Since $`det h \ne 0$, equations \eqref{Ke1} imply that $f^i$ do not depend on $t^1,t^2$.

Functions $f^1(u^1,u^2)$, $f^2(u^1,u^2)$ now satisfy the remaining equations \eqref{Ke2}
and also derivatives of equations \eqref{Ke2} with respect to $t^1,t^2$. The resulting nine homogeneous linear equations can be viewed as an algebraic system in four unknowns
$\frac{\partial f^i}{\partial u^k}$, $i,k = 1,2$.
Its properties depend on the values of six $4 \times 4$ minors
$D_j$, $j = 1,\dots,6$, formed by the rows corresponding to the three equations \eqref{Ke2} plus one of the six derivatives thereof.
These minors are
$$
D_1 = 8 (h_{11} h_{12,1} - h_{12} h_{11,1}) `det h, \quad
D_4 = 8 (h_{11} h_{12,2} - h_{12} h_{11,2}) `det h, \\
D_2 = 4 (h_{11} h_{22,1} - h_{22} h_{11,1}) `det h, \quad
D_5 = 4 (h_{11} h_{22,2} - h_{22} h_{11,2}) `det h, \\
D_3 = 8 (h_{12} h_{22,1} - h_{22} h_{12,1}) `det h, \quad
D_6 = 8 (h_{12} h_{22,2} - h_{22} h_{12,2}) `det h. \\
$$
If at least one of them is nonzero, then the algebraic system \eqref{Ke2} has only the zero solution, hence $f^1,f^2$ are constants and $\xi_3$ belongs to $\mathfrak K$.

If, conversely, $D_j = 0$ for all $j = 1,\dots,6$, then
$$
\numbered\label{Hc}
h_{ij} = H c_{ij}
$$
where $H(t^1,t^2)$ is a nonzero scalar function and $c_{ij}$ is a nonsingular symmetric constant matrix (meaning that coefficients $h_{ij}$ are constant multiples of each other).
Under \eqref{Hc}, system \eqref{Ke2} has a unique solution
$$
\xi_3 = -(c_{11} u^1 + c_{12} u^2) \frac{\partial}{\partial u^2}
 + (c_{12} u^1 + c_{22} u^2) \frac{\partial}{\partial u^1}
$$
modulo $\mathfrak K$ and up to a constant multiple.
\end{proof}

By Proposition~\ref{aKv} we have either $\xi_3 \in \mathfrak K$ or
\eqref{Hc}. In the latter case $C_\chi = 4 C_\rho$, $Q_\chi = 0$,
$Q_\gamma = 0$, contradicting assumption {\rm 3}.

\subsection{Vacuum case}

We consider Einstein's vacuum equations with cosmological constant
\begin{equation} \label{EE}
\mathbf{Ric}_{ij} - \frac{1}{2} \mathbf R \mathbf g_{ij} + \Lambda
\mathbf g_{ij} = 0
\end{equation}
where $\Lambda $ denotes the cosmological constant.

\begin{proposition}
\label{vacuum} Let $ C_{\gamma}^2 \neq 4 Q_{\gamma}. $ Let $p =
C_\rho$, let $q$ be an arbitrary invariant from the list  $C_\chi,\
Q_\chi,\ Q_\gamma$, such that $p,q$ are functionally independent (on
$\mathcal S$). Then knowing the values of the remaining two
invariants in terms of $p,q$ are sufficient for recovering a generic
metric $\bigmetric \in ~\mathfrak T$ in the vacuum case.
\end{proposition}

\begin{proof}
Using the Weyl--Lewis--Papapetrou parametrization, we consider the
system consisting of six Einstein's vacuum equations with
cosmological constant~\eqref{EE} and eight equations for $XC_\rho,
XC_\chi, XQ_\chi, XQ_\gamma, YC_\rho, YC_\chi, YQ_\chi, YQ_\gamma$.
This system of 14 equations is linear in second derivatives of the
four functions $f, s, r, w$. The number of second derivatives is ten
($f_{tt}$ and $f_{zz}$ are missing). By applying linear algebra we
obtain  four nontrivial linear dependences among the second order
invariants and components of Einstein's vacuum equations with
cosmological constant. Rewritten in invariant form, they are
\begin{equation} \nonumber
 YC_{\rho} = - 8 I_{1} ,  \\
XC_{\rho} = 8 (Q_{\gamma} - Q_{\chi}) + C_{\rho}(C_{\gamma} -
\frac{3}{4} C_{\rho} - 4 L ) , \\
\frac{1}{4} C_{\gamma}  I_{1}  XC_{\chi} - \frac{1}{2} I_{1}
XQ_{\gamma}  + \frac{1}{2} I_{2} YQ_{\gamma} -
\frac{1}{2} I_3  YC_{\chi} - I_4 YQ_{\chi} \\
\quad = I_{1} [I_4( C_{\gamma} -\frac{1}{2} C_{\rho} + 4\Lambda) -
\frac{1}{16}
C_{\gamma}C_{\rho}(\frac{3}{4} C_{\rho} + 4 \Lambda)], \\
\frac{1}{4} \varepsilon C_{\gamma}  I_{1}  YC_{\chi}  - \frac{1}{2}
\varepsilon I_{1} YQ_{\gamma}  - \frac{1}{2} I_2
XQ_{\gamma} + \frac{1}{2} I_3 XC_{\chi} + I_4  XQ_{\chi} \\
\quad = I_4((\frac{1}{2}  C_{\rho} + 4 \Lambda)( Q_{\gamma} - Q_{\chi}) + 2
I_3) - \frac{1}{8} I_3 C_{\rho}(\frac{3}{4} C_{\rho} + 4 \Lambda)
\end{equation}
where we denoted
\begin{equation} \nonumber
{I_1}^2 = \varepsilon [(Q_{\gamma} -
Q_{\chi})^2 -\frac{1}{4} C_{\rho}
(C_{\gamma} Q_{\chi} - (C_{\gamma} + \frac{1}{4} C_{\rho}) Q_{\gamma})],\\
I_2 =  -
(Q_{\gamma} + Q_{\chi}) + \frac{1}{2} C_{\gamma} (C_{\gamma} + \frac{1}{4} C_{\rho}) , \\
I_3 = \frac{1}{2}
 C_{\gamma} (Q_{\gamma} - Q_{\chi}) + \frac{1}{4} C_{\rho} Q_{\gamma}, \\
I_4 = \frac{1}{4} C_{\gamma}^2 - Q_{\gamma}, \\
C_{\rho} = 4 (C_{\chi} - C_{\gamma}), \\
\varepsilon = \pm 1
\end{equation}
(minus in elliptic case and plus in hyperbolic case).

The derivation of these relations is tedious but straightforward and
relies on inversion of a matrix with determinant $C_\rho I_4$. While
$C_\rho \ne 0$ is nonzero by assumptions (Section~\ref{gener}), $I_4
\ne 0$ is an additional genericity assumption.

To finish the proof, we express $Xp, Yp$, where $p = C_{\rho}$, from the
first and the second equation and $Xq, Yq$, where $q$ is one of $C_\chi,\
Q_\chi,\ Q_\gamma$, from the third and the fourth equation. By
Proposition~\ref{EP} this is sufficient for recovering the metric
$\bigmetric \in \mathfrak T$.
\end{proof}

For instance, if $q = Q_{\chi}$, then expressing $Xq, Yq$ is possible if and
only if $I_4 \neq 0$.

\begin{proposition} \label{psi2}
In the vacuum case, $`Re\Psi_2$ is of the first order.
\end{proposition}
\begin{proof}
On one hand the trace of $(\mathbf{Ric}|_{\mathcal S})_{ij}$ is
\begin{equation} \label{tr1}
g^{ij}(\mathbf{Ric}|_{\mathcal S})_{ij} = \mathbf R - 2 \Lambda
\end{equation}
 by \eqref{EE}. On the other hand the trace of \eqref{RSij} is
\begin{equation} \label{tr2}
g^{ij}(\mathbf{Ric}|_{\mathcal S})_{ij} = R - \frac12 C_\nu - \frac14
C_\rho + \frac12 C_\chi.
\end{equation}
 From \eqref{tr1}, \eqref{tr2} and by
\eqref{mathbf R} we have
\begin{equation} \label{Cnu}
C_\nu = -\frac1{2} C_\rho - 4 \Lambda.
\end{equation}
Thus, $C_\nu$ is a first order invariant in the vacuum case. The
trace of \eqref{EE} is
\begin{equation} \label{R}
\mathbf R = 4 \Lambda.
\end{equation}
Therefore, in the vacuum case we have
\begin{equation}
R = - \frac12 C_\chi.
\end{equation}
by \eqref{tr1}, \eqref{tr2}, \eqref{Cnu} and \eqref{R}.
 Then
\begin{equation} \nonumber `Re\Psi_2 = \frac1{8} C_\chi + \frac1{6}
\Lambda \end{equation}
 by \eqref{RePsi2}.
\end{proof}
The other Petrov invariant $\Psi_0\Psi_4$ is of second order even in
vacuum.

\section{Example}

\begin{example} \rm
We illustrate our results by classifying the Kerr--NUT--(anti)de Sitter
space-time, attributed to Pleba\'nski and Demia\'nski~\cite{P-D,S-K-M-H-H},
who derived it a special case of a seven-parameter family of Petrov type D solutions
with nonzero cosmological constant~$\Lambda$.
Denoting the nonignorable coordinates as $t_1,t_2$, the metric coefficients are
$$
h_{11} = \frac{P}{t_{1}^2 + t_{2}^2} -\frac{Q}{t_{1}^2 + t_{2}^2},
\quad
h_{12} = \frac{P t_{2}^2}{t_{1}^2 + t_{2}^2}
 + \frac{Q t_{1}^2}{t_{1}^2 + t_{2}^2}, \quad
h_{22} = \frac{P t_{2}^4}{t_{1}^2 + t_{2}^2}
 - \frac{Q t_{1}^4}{t_{1}^2 + t_{2}^2},
$$
and
$$
g_{11} =\frac{t_{1}^2 + t_{2}^2}{P}, \qquad g_{12} =  0, \qquad
g_{22} =  \frac{t_{1}^2 + t_{2}^2}{Q}
$$
where
$$
P = (A^2 - t_{1}^2)(1 + \frac{1}{3} \Lambda t_{1}^2) + 2 L t_{1}, \qquad
Q = (A^2 + t_{2}^2)(1 -\frac{1}{3} \Lambda t_{2}^2) - 2 M t_{2},
$$
$M, L, A, \Lambda$ being parameters of the metric.
While $M$ is the mass of the source, the physical meaning of $L,A$ is not immediately obvious. Griffiths and Podolsk\'y~\cite{G-P} gave the following identification
$$
A^2 =  (a^2 - l^2) \frac{(a^2 + 3 l^2) \Lambda + 3}{(a^2 - l^2) \Lambda + 3}, \\
L = l\left[\frac{1}{3} (a^2 - l^2)\Lambda
 + \frac{(a^2 + 3 l^2)\Lambda + 3}{(a^2 - l^2)\Lambda + 3}\right],
$$
where $a$ is the angular momentum and $l$ is the NUT parameter.

The four basic first-order invariants $C_\rho, C_\chi, Q_\chi, Q_\gamma$ can be routinely computed. Two of them turn out to be particularly simple, hence we choose them as the independent invariants $p,q$:
\begin{equation}
\label{pq}
p = Q_\gamma = -4 \left[\frac{ {\it L} t_{2} ( t_{2}^2 - 3
t_{1}^2)  - M t_{1} (3 t_{2}^2 - t_{1}^2 ) }{(t_1^2 + t_2^2)^3}
\right]^2,
\\
q = C_{\chi} = 4 \frac{ M t_{2} ( t_{2}^2 - 3 t_{1}^2) + {\it L} t_{1} (
3 t_{2}^2 - t_{1}^2) }{(t_1^2 + t_2^2)^3} - \frac{4}{3} \Lambda.
\end{equation}

Components of the invariant frame are easily seen to be
$$
X^{1} =  -\frac{2}{3} \frac{2 \Lambda t_{1}^3 - (\Lambda A^2 - 3) t_{1} - 3 L}
  {t_{1}^2 + t_{2}^2}, \\
X^{2} =   -\frac{2}{3} \frac{2 \Lambda t_{2}^3 + (\Lambda A^2 - 3) t_{2} + 3 M}
  {t_{1}^2 + t_{2}^2}, \\
Y^{1} =    \mathop{\rm sgn} Q \,\sqrt{\left|\frac{P}{Q} \right|} X^2, \\
Y^{2} =   -\mathop{\rm sgn} P \,\sqrt{ \left|\frac{Q}{P} \right|} X^1.
$$

Expressions for the other invariants needed for recognizing equivalence are quite large. They are schematically given in the following table:

\begin{center}
\renewcommand{\arraystretch}{1.3}
\begin{tabular}{c|cc}
Invariant & Numerator & Denominator \\ \hline
$C_{\rho}$ & $p_{{10,6,6}}$ & $(t_1^2 + t_2^2) PQ$ \\
$Q_{\chi}$ & $p_{{17,14,14}} $ & $(t_1^2 + t_2^2) ^6 PQ $ \\
${\it XQ_{\gamma}}$ & $p_{{3,3,3}}\,p_{{7,7,7}}$ & $(t_1^2 + t_2^2) ^{8} $  \\
${\it XC_{\chi}}$ & $p_{{7,7,7}} $ & $(t_1^2 + t_2^2) ^{5} $  \\
${\it YQ_{\gamma}}$ & $p_{{3,3,3}}\,p_{{11,8,8}} $ & $(t_1^2 + t_2^2) ^{8}{P}^{1/2}{Q}^{1/2} $  \\
${\it YC_{\chi}}$ & $p_{{11,8,8}} $  & $(t_1^2 + t_2^2) ^{5}{P}^{1/2}{Q}^{1/2}  $ \\
\end{tabular}
\end{center}
Here $p_{{n,n_1,n_2}}$ is a substitute for a polynomial $p(t_1,t_2)$
of total degree $n$, degree $n_1$ in the indeterminate
$t_1$ and degree $n_2$ in the indeterminate $t_2$ (e.g., the substitute for $P$ would be $p_{4,4,0}$).

Actual values of these invariants are suitable for storing in a database of exact solutions. Prior to that one can wish to express all of them in terms of $p,q$. This usually constitutes the main technical difficulty connected with equivalence analysis.
In our example a help comes from the simple relation
$$
(\frac{4}{3} \Lambda + C_\chi)^2  - 4 Q_\gamma =
16 \frac{M^2 + {\it L}^2}{(t_1^2 + t_2^2)^3} .
$$
Inserted back to the denominators in \eqref{pq}, this allows us to find the following cubic equations for $t_{1}$ and~$t_{2}$ in terms of invariants $p,q$ and parameters $\Lambda,M,L$:
$$4 t_{1}^3 - 3 I t_{1} + I_{+} = 0,
\\
4 t_{2}^3 - 3 I t_{2} - I_{-} = 0,
$$
where
$$
I = \left[
  16 \frac{M^2 + L^2} {(C_\chi + \frac{4}{3}\Lambda)^2 - 4 Q_\gamma}
\right]^{1/3} ,\qquad
I_{\pm} = 4 \frac{M (C_\chi + \frac{4}{3} \Lambda) \pm 2 L \sqrt{-Q_\gamma}}
{(C_\chi + \frac{4}{3} \Lambda)^2 - 4 Q_\gamma}.
$$

One easily checks that $I_4 \ne 0$ so that by Proposition~\ref{vacuum} the metric is characterizable by expressions for $C_\chi,Q_\gamma$ in terms of $C_\rho,Q_\chi$.
\end{example}

\begin{example} \rm
In the special case of $\Lambda = 0$ (Kerr--NUT solution)
we have $L = l$,
$P = -t_{1}^{2} + A^2 + 2\,Lt_{1}, Q = t_{2}^{2} + A^2 - 2\,Mt_{2}$ quadratic in $t_1,t_2$, respectively,
and the invariants simplify considerably:
\begin{center}
\renewcommand{\arraystretch}{1.3}
\begin{tabular}{c|cc}
Invariant & Numerator & Denominator \\ \hline
$C_{\rho}$ & $p_{{2,2,2}}$ & $(t_1^2 + t_2^2) PQ$ \\
$Q_{\chi}$ & $p_{{10,8,8}} $ & $(t_1^2 + t_2^2) ^6 PQ $ \\
${\it XQ_{\gamma}}$ & $p_{{3,3,3}}\,p_{{5,5,5}}$ & $(t_1^2 + t_2^2) ^{8} $  \\
${\it XC_{\chi}}$ & $p_{{5,5,5}} $ & $(t_1^2 + t_2^2) ^{5} $  \\
${\it YQ_{\gamma}}$ & $p_{{3,3,3}}\,p_{{7,6,6}} $ & $(t_1^2 + t_2^2) ^{8}{P}^{1/2}{Q}^{1/2} $  \\
${\it YC_{\chi}}$ & $p_{{7,6,6}} $  & $(t_1^2 + t_2^2) ^{5}{P}^{1/2}{Q}^{1/2}  $ \\
\end{tabular}
\end{center}

The same Kerr--NUT solution has been classified by \AA man and Karlhede in~\cite{A-Ka}. These authors start with the metric~\cite[II.A]{Ki}, given in the Newman--Penrose tetrad formalism. Since this metric is not in the form \eqref{gg}, the results cannot be compared immediately.
In the course of the Karlhede algorithm applied to this metric, 14 curvature invariants are computed. These are complex quantities, whose real and imaginary parts are to be considered separately.
This example confirms that a tailored procedure can lead to a significant reduction of the total number of computed invariants.

On the other hand, all the \AA man--Karlhede invariants already have (nearly) explicit expression in terms of two complex conjugated quantities $p,p^*$, which themselves are easily derivable from the unique complex curvature invariant of order zero. Thus the ``last step'' becomes rather trivial in this case. However, this may be attached to the particular way the Kerr--NUT metric (which is of Petrov type $D$) has been derived.
\end{example}

\section{Conclusions}

We have demonstrated that four first-order scalar differential invariants and two first-order invariant vector fields suffice to solve the local equivalence problem within the class $\mathfrak T$ of pseudo-Riemannian metrics with two commuting orthogonally transitive vector fields. Results are restricted to generic metrics in $\mathfrak T$ as identified in Sect.~\ref{gener}, leaving the special subclasses unexplored. The main open problems include: the description of the special (nongeneric) subclasses exempt from consideration in this paper; the classification problem within these classes; the comparison to the Karlhede algorithm; finding solutions to the Einstein equations in terms of first-order invariants (cf.~\cite{B-M} and references therein).

\section*{Acknowledgements}
Both authors gratefully acknowledge the support from M\v{S}MT under project MSM 4781305904.

\end{document}